\documentclass[aps,twocolumn,groupedaddress,showkeys,showpacs]{revtex4}
\usepackage[dvips]{graphicx}
\begin{document}

\title{Analytic study of the urn model for separation of sand}

\author{G.\ M.\ Shim}
\author{B.\ Y.\ Park}
\author{Hoyun Lee}
\affiliation{Department of Physics, Chungnam National University, 
Daejeon 305-764, R.\ O.\ Korea}

\date{\today}

\begin{abstract}
We present an analytic study of the urn model for separation of sand 
recently introduced by Lipowski and Droz (Phys.\ Rev.\ E {\bf 65}, 
031307 (2002)). 
We solve analytically the master equation and the first-passage problem. 
The analytic results confirm the numerical results obtained by 
Lipowski and Droz.
We find that the stationary probability distribution and the shortest 
one among the characteristic times are governed by the same {\it 
free energy}. We also analytically derive the form of the critical 
probability distribution on the critical line, 
which supports their results obtained by numerically calculating 
Binder cumulants (cond-mat/0201472).

\end{abstract}


\maketitle

\section{INTRODUCTION}

A granular system exhibits extremely rich phenomena, which has recently
attracted extensive studies. One of such interesting phenomena is the 
spatial separation of shaken sand \cite{Schlichting96}. Sand in a box 
separated into two equal parts by a wall that allows the transfer of 
sand through its narrow slit prefers to aggregate more in one side 
under certain conditions. 

Eggers explained the emergence of symmetry breaking using a hydrodynamic 
approach \cite{Eggers99}. The key idea is to introduce the effective 
temperature taking into account the inelastic collisions for granular 
material.

Lipowski and Droz proposed a dynamic model to explain the essence of
the phenomena \cite{Lipowski02a}. The model is a certain generalization 
of the Ehrenfest model \cite{Ehrenfest90}. Interestingly this model 
shows a spontaneous symmetry breaking in contrast to other 
generalizations of the Ehrenfest model. 
They derived the master equation and found in a numerical way 
the phase diagram that 
displays a rich structure like continuous and discontinuous transitions 
as well as a tricritical point. They also numerically solved the 
first-passage problem to find exponential or algebraic divergences.

Thanks to its simplicity, the model allows analytic approaches.
In this paper, we present the results of this analytic study to the 
master equation and the first-passage problem addressed by Lipowski 
and Droz. These not only confirm their numerical results but 
also give us some insights in the nature of the discontinuous transition 
in the stationary probability distribution.
We also analytically derive the form of the critical probability 
distribution on the critical line.

The paper is organized as follows. 
In Sec.~\ref{sec:II} we briefly review the model and its master equation. 
In Sec.~\ref{sec:III} we present the analytic solution of the master 
equation in the thermodynamic limit and the analytic expression of 
the stationary probability distribution. The form of the stationary
probability distribution on the critical line is also derived.
In Sec.~\ref{sec:IV} we analytically solve the first-passage problem.
Detailed analysis on the behavior of the characteristic times is given 
in Sec.~\ref{sec:V}. Section \ref{sec:VI} is devoted to conclusions 
and discussions.

\section{MODEL AND ITS MASTER EQUATION} \label{sec:II}

The model introduced by Lipowski and Droz \cite{Lipowski02a} is 
defined as follows. $N$ particles are distributed between two urns, 
and the number of particles in each urn is denoted as $M$ and $N-M$, 
respectively. 
At each time of updates one of the $N$ particles is randomly chosen. 
Let $n$ be a fraction of the total number of particles in the urn 
that the selected particles belongs to. 
With probability $\exp(-\frac{1}{T(n)})$ the selected particle moves 
to the other urn. $T(n)$ represents the effective temperature of an
urn with particles $nN$ that measures the thermal fluctuations 
of the urn.
Lipowski and Droz chose the temperature as $T(n) = T_0+\Delta(1-n)$. 

It is easy to derive the master equation for the probability
distribution $p(M,t)$ that there are $M$ particles in a given urn at 
time $t$ \cite{Lipowski02a}
\begin{eqnarray} 
   p(M,t+1) &=& F\bigl(\frac{N-M+1}{N}\bigr) p(M-1,t)
  \nonumber\\
            &&  + F\bigl(\frac{M+1}{N}\bigr) p(M+1,t)
  \nonumber\\
            &&  + \Bigl[  1-F\bigl(\frac{M}{N}\bigr)
                    -F\bigl(\frac{N-M}{N}\bigr) \Bigr] p(M,t)
\,,\label{eq:master}\end{eqnarray}
where $F(n) = n \exp(-\frac{1}{T(n)})$ measures the flux of 
particles leaving the given urn.
Here we introduced for convenience the notations $p(-1,t)=p(N+1,t)=0$.

The difference in the occupancy of the urns can be represented by 
the particle excess $\epsilon = \frac{M}{N}-\frac12$. The time 
evolution of the averaged particle excess
$e(t)=\langle\epsilon\rangle_t=\sum_M (\frac{M}{N}-\frac12) p(M,t)$
is governed by 
\begin{equation}\label{eq:averagedexcess}
   e(t+1) = e(t)+\frac{1}{N}\bigl\langle {\cal F}(\epsilon) 
                           \bigr\rangle_t
\,,\end{equation}
where ${\cal F}(\epsilon)= F(\frac12-\epsilon)-F(\frac12+\epsilon)$
measures the net flux of particles in the given urn. 
One conventionally takes the unit of time in such a way that there
is a single update per a particle on average. Therefore we scale
the time by $N$. Expanding Eq.~(\ref{eq:averagedexcess}) with respect 
to $\frac{1}{N}$, and using the mean--field approximation in evaluating
the average, we get
\begin{equation}\label{eq:differentialexcess}
  \frac{d}{dt}e(t) = {\cal F}\bigl(e(t)\bigr)
\,.\end{equation}
Note that the stationary solution of Eq.~(\ref{eq:differentialexcess}) 
is determined by zero points of ${\cal F}(\cdot)$. 
The stable stationary solutions are given by zero points of ${\cal F}(\cdot)$
with a negative slope, which we call as the stable fixed points while the
unstable ones are given by those with a positive slope, which we call as
the unstable fixed points.

\begin{figure}[tbp]
   \includegraphics[width=7cm]{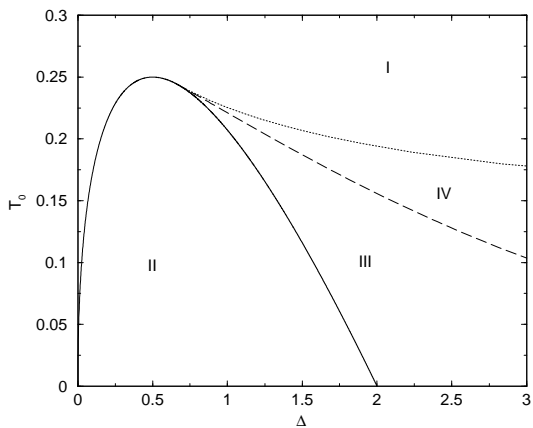} 
   \caption{\label{fig:phasediagram} Phase diagram of the urn model
            \cite{Lipowski02a}.
            The symmetric solution vanishes continuously on the solid line
            while the asymmetric one disappears discontinuously on the 
            dotted line. The transition of the behavior of the stationary
            probability distribution is denoted by the dashed line.}
\end{figure}

Detailed analysis on the existence of the stable stationary solutions of
Eq.~(\ref{eq:differentialexcess}) was done by Lipowski and Droz
\cite{Lipowski02a}. We here display their phase diagram in
Fig.~\ref{fig:phasediagram} to make our paper
as self-contained as possible. 
The stable symmetric solution ($\epsilon=0$) exists in region I, III, and
IV while the stable asymmetric solution ($\epsilon>0$) exists in region II,
III and IV.

\section{THE SOLUTION OF THE MASTER EQUATION} \label{sec:III}

We are mainly interested in investigating the properties of the 
infinite system.
Consider the thermodynamic limit $N \rightarrow \infty$ with
$\frac{M}{N}=\frac12+\epsilon$ being fixed. 
Representing the probability distribution by $\epsilon$ instead
of $M$, and expanding Eq.~(\ref{eq:master}) with respect to 
$\frac{1}{N}$, and keeping the terms up to the first order,
we arrive at the expression
\begin{eqnarray}
   p(\epsilon,t+1) &=& p(\epsilon,t)+\frac{1}{N}\Bigl[
                        \bigl(  F^\prime(\frac12+\epsilon)
                               +F^\prime(\frac12-\epsilon)
                        \bigr) p(\epsilon,t)
   \nonumber\\
                   &&   +\bigr(  F(\frac12+\epsilon)
                               -F(\frac12-\epsilon)
                        \bigr) \frac{\partial}{\partial \epsilon}
                               p(\epsilon,t)
                    \Bigr]  
\,.\label{eq:masterexpand}\end{eqnarray}

Scaling again the time by $N$, expanding Eq.~(\ref{eq:masterexpand}) 
with respect to $\frac{1}{N}$, and noting that the second term in the 
right-handed side can be combined into a total derivative with respect to
$\epsilon$, we finally obtain the partial differential equation
\begin{equation}\label{eq:continuousmaster}
    \frac{\partial}{\partial t}p(\epsilon,t)
   + \frac{\partial}{\partial \epsilon}\bigl[ 
         {\cal F}(\epsilon)p(\epsilon,t)
     \bigr] = 0 
\,.\end{equation}

\begin{figure}[tbp]
   \includegraphics[width=7cm]{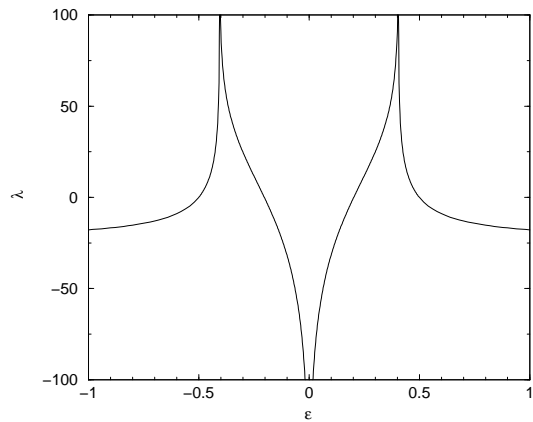} 
   \caption{\label{fig:map} $\lambda(\epsilon)$ for $\Delta=0.3, T_0=0.2$. 
            The mapping for $\epsilon>\frac12$ is analytically continuated.}
\end{figure}

Note that ${\cal F}(\cdot)$ is zero at a finite number of points.
The solution of Eq.~(\ref{eq:continuousmaster}) can be found in the 
intervals that do not include those points. At each interval, it would 
be convenient to introduce a new variable
\begin{equation}\label{eq:parametrization}
    \lambda(\epsilon) = \int_{\epsilon_0}^\epsilon \frac{dx}{{\cal F}(x)}
\,,\end{equation}
where $\epsilon_0$ is a certain point in the interval. 
Figure \ref{fig:map} shows a typical behavior of the mapping. 
We also displayed the map for $|\epsilon|>\frac12$ by analytic
continuation. This is necessary since the solution of 
Eq.~(\ref{eq:continuousmaster}) is of wave nature.
(See Eq.~(\ref{eq:wavesolution}) below.)
$\lambda$ increases as
$\epsilon$ approaches to the stable fixed points while it decreases as 
$\epsilon$ approaches to the unstable fixed points.
With the help of this parameterization and setting 
$R(\lambda,t)={\cal F}(\epsilon)p(\epsilon,t)$, 
Eq.~(\ref{eq:continuousmaster}) now takes the form
\begin{equation}\label{eq:wave}
    \frac{\partial}{\partial t}R(\lambda,t)
   +\frac{\partial}{\partial \lambda}R(\lambda,t) = 0
\,.\end{equation}
Note that Eq.~(\ref{eq:wave}) is in fact a half part of
the wave equations so that its solution is written as 
$R(\lambda,t) = f(\lambda-t)$ with $f(\cdot)$ being an arbitrary
differentiable function. It represents a wave that moves to the direction
of increasing $\lambda$ as time evolves, which means that the system
moves to stable fixed points. 
The solution of the original partial differential equation
(\ref{eq:continuousmaster}) now reads
\begin{equation}\label{eq:wavesolution}
    p(\epsilon,t) = \frac{ f\bigl(\lambda(\epsilon)-t\bigr) }{
                           {\cal F}(\epsilon) } 
\,.\end{equation}

From the initial probability distribution $p_0(\epsilon) = p(\epsilon,0)$
we can determine the function $f(\cdot)$. So we get 
\begin{equation}\label{eq:solutiondistrib} 
   p(\epsilon,t) = \frac{{\cal F}(\epsilon_t)}{{\cal F}(\epsilon)}
                   p_0(\epsilon_t)
\,,\end{equation}
where $\epsilon_t$ is given by the relation
\begin{equation}\label{eq:inversemap}
    \lambda(\epsilon_t) = \lambda(\epsilon)-t
\,.\end{equation}
Here it should be understood that $\epsilon_t$ is to be chosen in
the same interval where $\epsilon$ belongs to. 
Furthermore $p_0(\epsilon)=0$ for $|\epsilon|>\frac12$ is assumed since
$\epsilon_t$ in Eq.~(\ref{eq:solutiondistrib}) can be larger 
(or smaller) than $\frac12$ (-$\frac12$). This happens because of the
nature of the wave solution.
Using the mapping from $\epsilon$ to $\epsilon_t$, it is straightforward to
show that the total probability is conserved:
\begin{equation}
       \int_{-\frac12}^\frac12 p(\epsilon,t)d\epsilon
     = \int_{-\frac12}^\frac12 p_0(\epsilon_t)d\epsilon_t
     = 1
\,.\end{equation} 

The shape of the probability distribution is distorted by a ratio 
${\cal F}(\epsilon_t)/{\cal F}(\epsilon)$ so that it accumulates
at the nearest stable fixed points.
In Eq.~(\ref{eq:solutiondistrib}), the ratio approaches zero as time 
evolves unless $\epsilon$ is on a stable fixed point. 
As a consequence, in the long time limit $t \rightarrow \infty$ 
the probability distribution becomes a sum of delta peaks at the stable
fixed points  $\epsilon_i$
\begin{equation}\label{eq:probdistinfty}
    p(\epsilon,\infty) = \sum_i p_i \delta(\epsilon-\epsilon_i)
\,,\end{equation}
where $p_i$ are the sum of the initial probabilities 
in two intervals adjacent to its point $\epsilon_i$.
We would like to point out that the system is not ergodic and its dynamical
phase space is decomposed into disconnected sectors. Each sector is
associated with a stable fixed point and is separated by the unstable fixed
points.

The fixed point condition ${\cal F}(\epsilon)=0$ is equivalent to Eq.~(4)
in Ref.~\cite{Lipowski02a}, where Lipowski and Droz analyzed in detail the
condition and their results are summarized in the phase diagram (See
Fig.~\ref{fig:phasediagram}.). 
We would like to mention that 
in regions III and IV in Fig.~\ref{fig:phasediagram} both the
symmetric and the asymmetric solutions exist together. In fact, either
solution can be realized by choosing an appropriate initial configuration.
Lipowski and Droz distinguished regions III and IV according to the 
different behaviors of the stationary probability distribution
in their numerical process of taking the limit $N \rightarrow \infty$. 
In region III there appear two delta peaks for the
asymmetric solutions while in region IV there appears only the central delta 
peak for the symmetric solution. It is contradictory to our result 
Eq.~(\ref{eq:probdistinfty}) where any
delta peaks for the stable fixed points can appear depending on the initial
configurations.

To resolve this contradiction and understand the nature of the transition
between regions III and IV, we take another limit in the master
equation (\ref{eq:master}), namely take the long time limit $t \rightarrow
\infty$ before we take the limit $N \rightarrow \infty$. This limit may not
properly reflect the properties of the infinite system. 
Since the infinite system is not ergodic as we showed above,
changing the order of taking limits $N \rightarrow \infty$ and
$t \rightarrow \infty$ may not yield the same result. 
Anyway it seems that in their simulations about the stationary probability 
distribution, Lipowski and Droz took the limit $t \rightarrow \infty$ 
for a finite system size $N$, and then extrapolating the results 
to $N \rightarrow \infty$.

Let's first take the long time limit of $t \rightarrow \infty$ in 
Eq.~(\ref{eq:master}). In this limit we may drop off the time dependence in
the probability distribution, which now takes the form
\begin{eqnarray}
   p(N) &=& F\bigl(\frac1N\bigr)p(N-1)+\bigl(1-F(1)\bigr)p(N) 
 \nonumber\\
   p(M) &=& F\bigl(\frac{N-M+1}{N}\bigr)p(M-1)
 \nonumber\\
        &&    +F\bigl(\frac{M+1}{N}\bigr)p(M+1)
 \nonumber\\
        &&    +\Bigl[ 1-F\bigl(\frac{M}{N}\bigr)-
                      F\bigl(\frac{N-M}{N}\bigr)\Bigr]p(M)
 \nonumber\\
         && \mbox{for $M=N-1,\ldots,2,1$}
 \nonumber\\
   p(0) &=& F\bigl(\frac1N\bigr)p(1)+\bigl(1-F(1)\bigr)p(0)
\,.\label{eq:stationaryprob}\end{eqnarray}
The first equation in Eq.~(\ref{eq:stationaryprob}) allows us to rewrite 
$p(N)$ in terms of $p(N-1)$, which in turn allows to rewrite $p(N-1)$ 
in terms of $p(N-2)$, and so on. Therefore we find
\begin{eqnarray}
    p(M) &=& \frac{ F\bigl(\frac{N-M+1}{N}\bigr) }{
                    F\bigl(\frac{M}{N}\bigr)} p(M-1)
  \nonumber\\
         &=& p(0) \prod_{i=1}^M \frac{ F\bigl(\frac{N-i+1}{N}\bigr) }{
                   F\bigl(\frac{i}{N}\bigr)}
\label{eq:stationaryrecursive}\end{eqnarray}
for $M=N,\ldots,2,1$. $p(0)$ appears as an overall factor to normalize
the probabilities so that we get
\begin{equation}
   p(0) = \Bigl[ 1+\sum_{M=1}^N \prod_{i=1}^M 
            \frac{F\bigl(\frac{N-i+1}{N}\bigr)}{F\bigl(\frac{i}{N}\bigr)}
          \Bigr]^{-1}
\,.\end{equation}
Now let's take the limit $N \rightarrow \infty$.
With $\frac{M}{N} = \frac12+\epsilon$, and 
$\frac{i}{N}= \frac12+x$, and scaling the probability distribution by
$N$, the stationary probability distribution for large $N$ now becomes
\begin{equation}\label{eq:statcontinprob}
   p_s(\epsilon) \approx \frac{ e^{NG(\epsilon)} }{
                 \int_{-\frac12}^\frac12 dx \, e^{NG(x)} }
\,,\end{equation}
where 
\begin{equation}
    G(\epsilon) = \int_{-\frac12}^\epsilon dx \,
                  \bigl[ \log F(\frac12-x) - \log F(\frac12+x) \bigr]
\,.\end{equation}

In the limit $N \rightarrow \infty$, the main contribution to the
stationary probability distribution comes only from the maximum of
$G(\cdot)$, and it becomes delta peaks. 
The maximum of $G(\epsilon)$ occurs when
\begin{equation}
    G^\prime(\epsilon) =  \log F(\frac12-\epsilon) 
                         -\log F(\frac12+\epsilon) 
                       = 0
\,.\end{equation}
Since $G^\prime(\epsilon)$ is the difference of logarithms of 
$ F(\frac12-\epsilon) $ and $ F(\frac12+\epsilon) $, 
both $G(\cdot)$ and ${\cal F}(\cdot)$ share the similar qualitative 
properties. For example, the maximum of both functions occurs at 
the same stable fixed points. 
Note that in region II only the two asymmetric solutions 
with positive and negative particle excesses are stable,  and have the
same maximum  while in region I only the symmetric solution is stable. 
Therefore the stationary probability distribution has the double peaks 
in region II and  only the central peak in region I. 
In region III and IV both the
symmetric and the asymmetric solutions are stable so that the maximum
of $G(\epsilon)$ should be determined by comparing its values at the
stable fixed points. The crossover of the maximum point occurs when
both values coincide. This implies that the transition between
the double peaks and the central single peak in the probability
distribution is determined by the condition
$\Delta G = G(\epsilon_a)-G(0) = 0$, where $\epsilon_a$ is the nonzero
stable fixed point. This condition yields a line that separate two
regions III and IV.

It is very interesting to see that $-G(\cdot)$ resembles the free
energy of the equilibrium systems, and the transition between two
{\it phases} is determined by the condition that the free energies of
both phases are equal. Furthermore a certain characteristic time
behaves differently in two phases, as Lipowski and Droz numerically
found \cite{Lipowski02a}. We will show an analytic relation between
them in next section.

\section{CHARACTERISTIC TIMES}\label{sec:IV}

Lipowski and Droz defined an averaged first-passage time $\tau(M)$
needed for a configuration with $M$ particles in an urn (and $N-M$
particles in the other urn) to reach the symmetric configuration
($M=\frac{N}{2}$) \cite{Lipowski02a}. 
They obtained the relations among the averaged characteristic 
times from the dynamical rules as
\begin{eqnarray}
   && \hspace{-2em}\tau(M) = F\bigl(\frac{M}{N}\bigr)\bigl[\tau(M-1)+1\bigr]
   \nonumber\\
   && \hspace{1em}  +F\bigl(\frac{N-M}{N}\bigr)\bigl[\tau(M+1)+1\bigr]
   \nonumber\\
   && \hspace{1em} +\Bigl[1-F\bigl(\frac{M}{N}\bigr)
                    -F\bigl(\frac{N-M}{N}\bigr)\Bigr]
                \bigl[\tau(M)+1\bigr]
\label{eq:chartimerelation}\end{eqnarray}
for $M=N, N-1, \ldots, \frac{N}{2}+1$. Here it is understood that the
term associated with $\tau(N+1)$ does not appear. (In fact, its
coefficient $F(0)$ vanishes.) By definition of the characteristic
times, $\tau(\frac{N}{2})=0$ and $\tau(N-M) = \tau(M)$.

Defining the difference of successive characteristic times as 
$\Delta \tau(M)=\tau(M)-\tau(M-1)$, Eq.~(\ref{eq:chartimerelation}) can
be rewritten as
\begin{equation}
   \Delta \tau(M) = \frac{1}{F\bigl(\frac{M}{N}\bigr)}\Bigl[
                    1+F\bigl(\frac{N-M}{N}\bigr)\Delta\tau(M+1) \Bigr]
\,.\end{equation}
By applying this relation repeatedly until
$\Delta\tau(N)=\frac{1}{F(1)}$ is reached, we get the expression
\begin{equation}\label{eq:Dtaurelation}
   \Delta\tau(M) = \frac{1}{F\bigl(\frac{M}{N}\bigr)}\Bigl[
                   1+\sum_{i=1}^{N-M} \prod_{j=1}^i 
                    \frac{ F\bigl(\frac{N-M-j+1}{N}\bigr) }{
                            F\bigl(\frac{M+j}{N}\bigr) }
                  \Bigr]
\,.\end{equation}
Since $\tau(\frac{N}{2})=0$ by definition, we immediately get
$\tau(\frac{N}{2}+1) = \Delta\tau(\frac{N}{2}+1)$, which is given
by Eq.~(\ref{eq:Dtaurelation}) with $M=\frac{N}{2}+1$. 
By successively adding $\Delta\tau(M)$, we get the general expression 
for $\tau(M)$ for $M=N,N-1,\ldots,\frac{N}{2}+1$:
\begin{equation}
   \tau(M) = \sum_{k=\frac{N}{2}+1}^M \frac{1}{F\bigl(\frac{k}{N}\bigr)}
                 \Bigl[1+\sum_{i=1}^{N-k} \prod_{j=1}^i 
                    \frac{ F\bigl(\frac{N-M-j+1}{N}\bigr) }{ 
                            F\bigl(\frac{M+j}{N}\bigr) }
                  \Bigr]
\,.\end{equation}
                  
We are mainly interested in the behavior of $\tau(M)$ as $N$ increases.
Again we scale the characteristic times by $N$. 
With $\frac{M}{N}=\frac12+\epsilon$ being fixed, and introducing
$\frac{k}{N}=\frac12+x, \frac{i}{N}=y-x, \frac{j}{N}=z-x$, 
the summations for large $N$ can be replaced by integrations so that
the characteristic times for $\epsilon>0$ takes the form
\begin{equation}
   \tau(\epsilon) \approx N \int_0^\epsilon dx \int_x^\frac12 dy \,
                          e^{NH(x,y)}
\end{equation}
with $H(x,y) = G(y)-G(x)$. 
The longest characteristic time $\tau(N)$ corresponds to
$\tau(\epsilon=\frac12)$.  The shortest one $\tau(\frac{N}{2}+1)$ 
corresponding to $\tau(\epsilon=0$) is, in general, smaller than 
$\tau(\epsilon)$ with positive $\epsilon$ 
by an order of magnitude. It is necessary to deal with it separately.
We get
\begin{equation}
   \tau(\frac{N}{2}+1) \approx  \int_0^\frac12 dy \,
                          e^{NH(0,y)}
\,.\end{equation}

Since $H(0,y)=G(y)-G(0)$, both the shortest characteristic time
$\tau(\frac{N}{2}+1)$ and the stationary probability distribution 
$p_s(\epsilon)$ have essentially the same functional dependence on
$G(\cdot)$. Therefore it is not surprising that behaviors of both 
quantities for large $N$ are closely related. 
However it is not clear why they are.

\section{ANALYSIS OF THE CHARACTERISTIC TIMES}\label{sec:V}

We first consider the behavior of $\tau(\frac{N}{2}+1)$, the shortest
one among the characteristic times. For large $N$, the main
contribution to $\tau(\frac{N}{2}+1)$ comes from the maximum point
$y_m$ of $H(0,y)=G(y)-G(0)$, or $G(y)$, 
which was dealt in Sec.~\ref{sec:III}. 

In region I and IV the maximum occurs at $y_m=0$ corresponding to the
symmetric configuration, while in region II and III it occurs at 
$y_m>0$ corresponding to the asymmetric configurations. 
Therefore the maximum is zero in region I and IV while it is positive 
in region II and III.

When $y_m=0$, we may expand $H(0,y)$ around $y=0$ to get
\begin{eqnarray}
    H(0,y) &\approx& -2\Bigl[1-\frac{\Delta/2}{(T_0+\Delta/2)^2}\Bigr] y^2
  \nonumber\\
           && -\frac43\Bigl[1-\frac{3(\Delta/2)^2}{(T_0+\Delta/2)^4}
                      \Bigr] y^4 + \ldots
\,.\label{eq:Hyexpand0}\end{eqnarray}
Therefore it yields $\tau(\frac{N}{2}+1) \sim N^{-\frac12}$
as long as the coefficient of the first term in
Eq.~(\ref{eq:Hyexpand0}) is negative. This is the case in region I and
IV. The coefficient vanishes when 
$T_0 = \sqrt{\frac{\Delta}{2}}-\frac{\Delta}{2}$, 
which corresponds to the critical line found by Lipowski and Droz
\cite{Lipowski02a}. On this line, we get 
\begin{equation}\label{eq:scalingbehavior}
   H(0,y) \approx  -\frac43\bigl(1-\frac32\Delta\bigr) y^4
                   -\frac{32}{15}\bigl(1-\frac54\Delta^2\bigr)y^6
                   +\ldots
\,,\end{equation}
which yields $\tau(\frac{N}{2}+1) \sim N^{-\frac14}$ for
$\Delta < \frac23$, and $\tau(\frac{N}{2}+1) \sim N^{-\frac16}$ at
$\Delta = \frac23$ corresponding to the tricritical point.

When $y_m>0$, the maximum is positive so that 
$\tau(\frac{N}{2}+1) \sim N^{-\frac12}e^{\alpha N}$ (with $\alpha$ being
a positive constant), that is, it diverges exponentially.

Now let's investigate the behavior of the longest characteristic time
$\tau(N)$ or $\tau(\epsilon=1)$. Again the main contribution comes
from the maximum point of $H(x,y)$ in region restricted by three lines
$y=x, x=0, y=\frac12$. Note that $y \ge x$ in the region.
Interestingly the maximum point $(x_m,y_m)$ is closely related with 
the fixed points of ${\cal F}(\epsilon)$. 

\begin{figure}[tbp]
   \includegraphics[width=7cm]{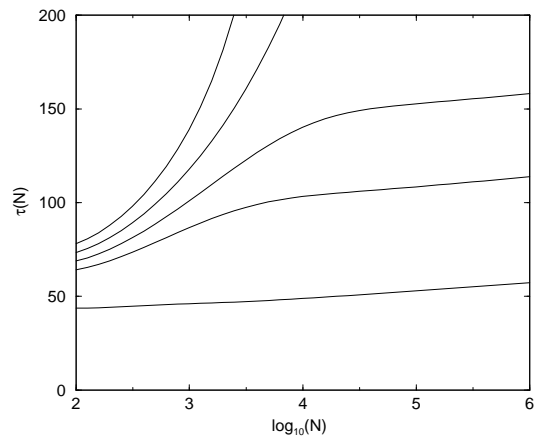} 
   \caption{\label{fig:tauNlogN} characteristic time  $\tau(N)$ as a 
            function of $N$ for $T_0=0.2$ and $\Delta=1.72, 1.744067, 
            1.77,1.8,2.0$ (from the top).}
\end{figure}

In region I, $\epsilon=0$ is only the fixed point so that $x_m=y_m=0$, 
and the maximum is zero. Expanding $H(x,y)$ about this point, we get
\begin{equation}\label{eq:HxyExpand0}
  H(x,y) \approx -2\Bigl[1-\frac{\Delta/2}{(T_0+\Delta/2)^2}\Bigr]
                   (y^2-x^2)+\ldots
\,.\end{equation}
The coefficient of the leading term in Eq.~(\ref{eq:HxyExpand0}) is
negative only if $T_0 > \sqrt{\frac{\Delta}{2}}-\frac{\Delta}{2}$, that
is, above the critical line. Changing the variables $x, y$ to the polar
coordinates $r, \theta$ and scaling the radial coordinate $r$ by
$\sqrt{N}$, we arrive at 
\begin{eqnarray}
  \tau(N) &\approx& \int_0^{c\sqrt{N}} dr \, r
                  \int_{\frac{\pi}{4}}^{\frac{\pi}{2}} d\theta \, 
                  \exp\Bigl[-2\Bigl(1-\frac{\Delta/2}{(T_0+\Delta/2)^2}
                              \Bigr) 
               \nonumber \\
          && \hspace{9em}\times\, r^2\cos(2\theta)\Bigr]
\,.\end{eqnarray}
Here $c$ is a constant that gives the upper bound of the integration 
over $r$. The contribution from the neighborhood of $\theta=\frac{\pi}{4}$
yields a logarithmic divergence. Fig.~\ref{fig:tauNlogN} shows a typical 
behavior of characteristic time $\tau(N)$ as a function of $N$ 
for several values of $\Delta$ with $T_0=0.2$. The first two uppermost lines
stand for $\tau(N)$ in region II, and the others represent that in region
I. We conclude that in region I $\tau(N)$ diverges logarithmically 
as $N$ increases.

As we see in Eq.~(\ref{eq:HxyExpand0}), the leading term vanishes on
the critical line. On this line we need to expand more. So
\begin{eqnarray}
   H(x,y) &\approx&  -\frac43\bigl(1-\frac32\Delta\bigr) (y^4-x^4)
       \nonumber\\
          &&         -\frac{32}{15}\bigl(1-\frac54\Delta^2\bigr)(y^6-x^6)
                      +\ldots
\,.\end{eqnarray}
Again changing the variables $x, y$ to the polar coordinates and scaling
the radial coordinate appropriately (by $N^\frac14$ or $N^\frac16$),
we conclude that $\tau(N)$ diverges algebraically as $N^\frac12$ on
the critical line and as $N^\frac23$ at the tricritical point.

In regions II, III, and IV 
there appear many fixed points among which we can always find one 
with $y_m>x_m$ and $G(y_m)>G(x_m)$. There, the maximum
$H(x_m,y_m)$ is positive, and $\tau(N)$ diverges exponentially
as $N$ increases. We would like to point out that the situation is
different from that of $\tau(\frac{N}{2}+1)$.
In fact, it corresponds to the case with $x_m$ being fixed to zero.

Finally let's consider the behavior of $\tau(N)$ on the line
separating two regions I and IV. As we approach this line from region
IV, the nonzero fixed points merge to disappear at
$\epsilon=\epsilon_1>0$. That is, we can find a positive $\epsilon_1$
such that ${\cal F}(\epsilon_1) = {\cal F}^\prime(\epsilon_1) = 0$.
On this line the maximum point is given by $x_m=y_m=\epsilon_1$ so that 
the maximum $H(x_m,y_m)$ is zero. 
Expanding $H(x,y)$ around this point, we get
\begin{equation}
     H(x,y) \approx -\frac16 G^{\prime\prime\prime}(\epsilon_1)
                    \bigl((y-\epsilon_1)^3-(x-\epsilon_1)^3\bigr)
                    +\ldots
\,.\end{equation}
(Here we don't write down $ G^{\prime\prime\prime}(\epsilon_1)$
explicitly since it is not important as far as it is positive.) 
Note that the leading order is the third instead of the fourth 
as in case of the critical line. 
The reason is that $G(\epsilon)$ is not symmetric about 
$\epsilon = \epsilon_1$ while it is symmetric about $\epsilon = 0$.
Consequently $\tau(N)$ on this line diverges algebraically as
$N^\frac13$.

\section{CONCLUSIONS}\label{sec:VI}

We analytically investigate the urn model introduced by Lipowski and
Droz \cite{Lipowski02a}. We exactly solve the master equation of the
model in the thermodynamic limit and find how the probability
distribution evolves. 
In the long time limit, the probability distribution becomes 
delta peaks only at the stable fixed points. 
In fact the ergodicity of the dynamics is broken so that the dynamical
phase space is decomposed into disconnected sectors separated by the
unstable fixed points. The strength of a delta peak is equal to the
sum of initial probabilities in the disconnected sector it belongs to.

We also solve exactly the stationary probability distribution where we
take the long time limit before we take thermodynamic limit.
Regardless of the initial probability distribution it shows double
peaks or a single central peak depending on the parameters of the
system. The final formula of the stationary probability distribution 
resembles that of the equilibrium systems, where the transition from 
the doubles peaks to the single peak is determined by the condition 
that {\it free energies} of two phases become equal.

Recently Lipowski and Droz \cite{Lipowski02b}
numerically calculated  Binder cumulants
of the urn model to find that the critical probability distribution
has the form $p(x) \sim e^{-x^4}$ on the critical line, and
$p(x) \sim e^{-x^6}$ on the tricritical point, where $x$ is the rescaled
order parameter proportional to the particle excess $\epsilon$.
As we showed, $G(\epsilon) \sim H(0,\epsilon)$ and
its behavior on the critical line (including the tricritical point)
is given by Eq.~(\ref{eq:scalingbehavior}). Therefore the critical
probability distribution has the form $p(\epsilon) \sim \exp[
-\frac43(1-\frac32\Delta)\epsilon^4]$ on the critical line,
and $p(\epsilon) \sim \exp[-\frac{128}{135}\epsilon^6]$ on the
tricritical point.
Our analytic result supports their numerical result.

The first-passage problem is analytically solved. 
Interestingly both the shortest characteristic time 
and the stationary probability distribution are governed by the
same {\it free energy}. 
Therefore the behavior of the shortest characteristic
time and the properties of the stationary probability distribution are
closely related.
The analytic results on the behavior of the characteristic times 
support the numerical results of Lipowski and Droz \cite{Lipowski02a}.

Finally it would be very interesting to understand why both the stationary 
probability distribution and the shortest characteristic time are 
governed by the same {\it free energy}.
It would also be of interest to extend our analytic study to many-urn 
models and other types of urn models.


\end{document}